\documentclass[aps,pra,preprint,groupedaddress,showpacs]{revtex4}
\begin{document}
\title{Non-Empty Quantum Dot as a Spin-Entangler}
\author{Chih-Lung Chou}
\email{choucl@cycu.edu.tw}
\affiliation{Department of Physics, Chung-Yuan Christian University,
Taoyuan, Taiwan 32023}
\date{\today}

\begin{abstract}
We consider a three-port single-level quantum dot system with one
input and two output leads. Instead of considering an empty dot,
we study the situations that two input electrons co-tunnel through
the quantum dot occupied by one or two dot electrons. We show that
electron entanglement can be generated via the co-tunneling
processes when the dot is occupied by two electrons, yielding
non-local spin-singlet states at the output leads. When the dot is
occupied by a single electron, net spin-singlet final states could
be generated by injecting polarized electrons to the dot system.
When the input electrons are unpolarized, we show that by
carefully arranging model parameters non-local spin-triplet
electrons can also be obtained at the output leads if the
dot-electron spin remains unchanged in the final state.
\end{abstract}

%\keywords{}
\maketitle

\section{Introduction}
The quantum theory of computation and information has been rapidly
developed in the past years on the basis of quantum mechanics. The
quantum computation and communication requires quantum operations
(quantum gates) on quantum bits (qubits). Unlike the classical bit
of information that always takes either $"0"$ or $"1"$ value, the
qubit is a two-level quantum system which can be prepared in
arbitrary superpositions of its two states, usually denoted as
$|0>$ and $|1>$. The principle of superposition also applies to
input and out registers of quantum circuits. The principle of
superposition has no correspondence in classical information
theory and leads to the so-called quantum parallelism
\cite{QParallel} and other novel properties in quantum computation
theory \cite{QParallOthers}. Up to date, several quantum systems
have been considered as qubits and quantum gates. For examples,
nuclear magnetic resonance quantum computing \cite{NMR}, neutral
atoms \cite{neutralAtom}, low-capacitance Josephson junctions
\cite{JJQubit}, trapped ions \cite{ionsQubit}, cavity QED schemes
\cite{QEDscheme}, and the spin states of coupled semiconductor
quantum dots \cite{QdotsGates} etc., are proposed or have been
used to implement qubits and universal quantum gates for quantum
computation .

In addition to the principle of superposition, the nonlocal nature
of entangled quantum states also play important roles in quantum
computation and communication. The Einstein-Podolsky-Rosen (EPR)
pairs \cite{EPR} are essential in many striking phenomena such as
quantum teleportation \cite{teleport}, dense coding
\cite{denseCode}, secure quantum communication
\cite{secureCommunicate}, and the tests of the violation of Bell's
inequality \cite{Bell}. These interesting phenomena have been
tested in experimental demonstrations which used entangled photons
as EPR pairs. Although such tests exist for photons, they have not
been tested for electrons since it is difficult to produce and
detect entangled electrons in solid-state environment. Recently,
several proposals have been raised to generate \cite{Oliver,
GateQwire, 2dotsEntangle} and detect \cite{ProbEntangle} entangled
electrons . There are several main motivations for these
proposals. First, electron spins in semiconductor environment have
been demonstrated to have long dephasing times approaching
microseconds \cite{spindephase}. Since the qubits defined as
electron spins are mobile, they can be good candidates for
implementing quantum communication in solid-state environment.
Second, the EPR electrons can be spatially separated from each
other thus the control of nonlocality can be implemented. Third,
solid-state qubits may be integrated into large quantum networks
which are required for the realization of quantum computers.

Among the proposals for generation of electron entanglement, many
of them propose quantum dots as the spin-entanglers \cite{Oliver,
2dotsEntangle}. Recher {\it et al.} \cite{2dotsEntangle} suggest
using two coupled quantum dots that are coupled to one s-wave
superconductor input lead and two normal Fermi output leads as a
electron spin-entangler. Two electrons from the superconductor
tunnel coherently into different output leads via different
quantum dots by the Andreev process in the Coulomb blockade regime
with the presence of voltage bias. Oliver {\it et al.}
\cite{Oliver} suggest using an empty single-level quantum dot with
one input and two output leads as the spin-entangler. The leads
are nondegerate and of narrow widths in energy so that they act as
energy filters for the input and output electrons. The leads are
arranged to suppress any single-electron tunneling process. Only
two input electrons can co-tunnel from the input lead into
different output leads via the dot. In this scheme, the double
occupancy of the dot will incur an on-site Coulomb interaction
which then mediates electron entanglement. Therefore, two input
electrons with opposite spins will be in the spin-singlet state at
the output leads after co-tunneling through the dot.

In this paper, we showed that the quantum dot needs not to be an
empty one to become a spin-entangler as in the scheme proposed by
Oliver {\it et al.}\cite{Oliver}.  In contrast, the quantum dot
can be occupied either by a single electron or by two electrons
with different spins. When there is one electron occupying the
dot, it is possible to obtain a net spin-triplet amplitude at the
output leads if the dot electron does not flip its spin after the
co-tunneling processes. When there are two electrons with
different spins occupying the dot, a net spin-singlet state is
obtained after co-tunneling two input electrons. The other final
states such as the spin-triplet and the same-spin states at the
output leads destructively interfere in this case. It is noted
that the on-site Coulomb interaction still plays the role of
entanglement mediator as usual. Therefore when the Coulomb
interaction is turned off, all possible final states destructively
interfere thus give vanishing transition amplitudes.

\section{The quantum dot system}
\begin{figure}
\vspace*{1.3in}\includegraphics{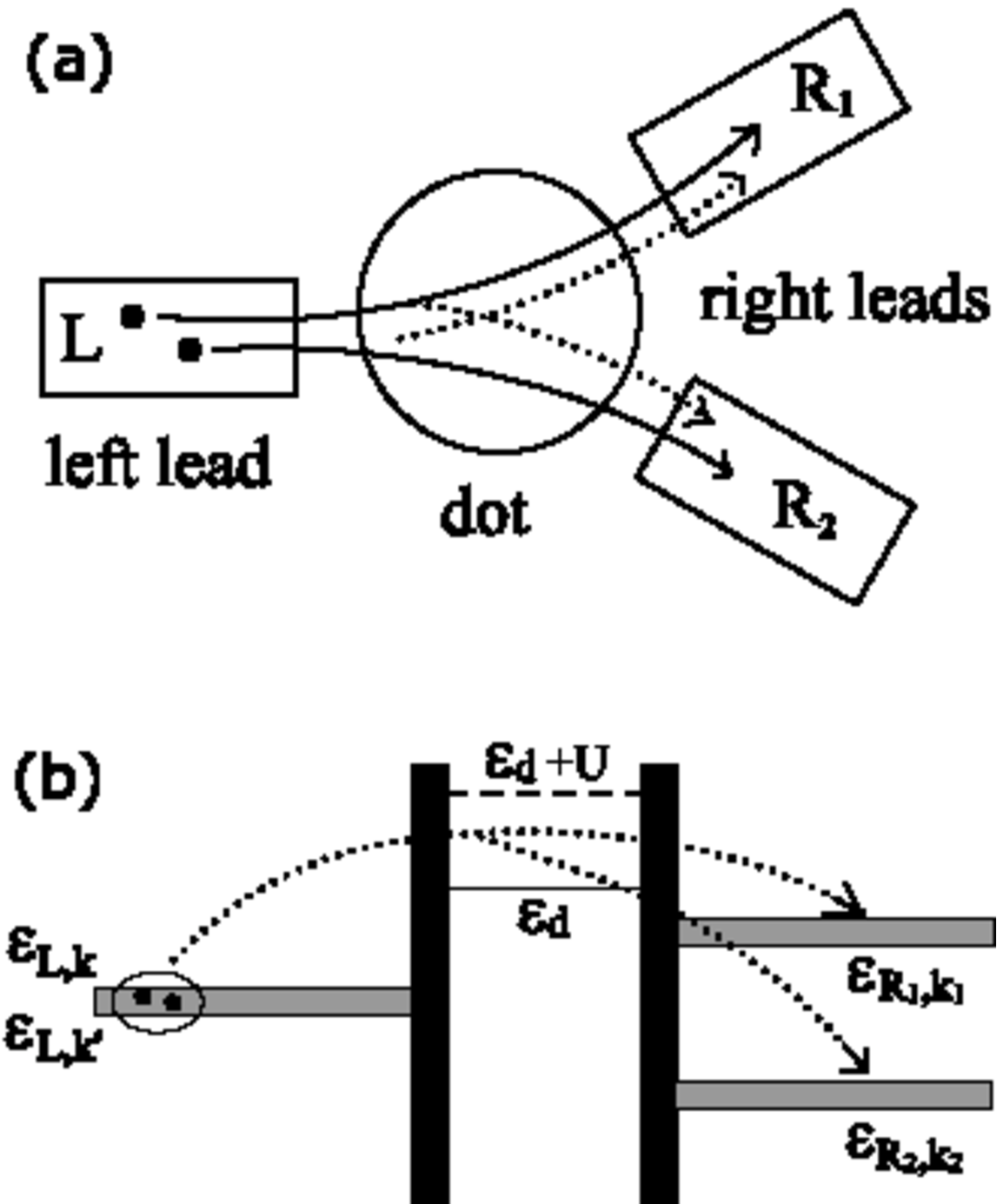}\vspace{2.0in} \caption{(a)The
quantum dot system with three leads. (b) The energy band diagram
of the quantum dot system.} \label{fig:DotSystem}
\end{figure}

The arrangement of the quantum dot system and its energy band
diagram are given in Ref.\cite{Oliver} and also in
Fig.\ref{fig:DotSystem}. Through out this paper we assume that the
quantum dot has only a single spin-degenerate energy level, thus
no single electron excitation inside the dot is allowed. According
to Pauli's principle of exclusion, the dot can accommodate at most
two electrons with different spins. Same-spin electrons will not
co-exist within the dot. An on-site Coulomb interaction is assumed
to incur when two different-spin electrons occupy the dot. In
general, the quantum dot system is described by the Anderson
Hamiltonian $\hat{H}$ as follows:
\begin{eqnarray}
\hat{H}&=&\hat{H_0} + \hat{V}, \nonumber \\
\hat{H_0}&=&\sum_{\eta, k, \sigma}\varepsilon_{\eta,
k}\hat{a}_{\eta, k, \sigma}^{+}\hat{a}_{\eta, k,
\sigma}+\sum_{\sigma}\varepsilon_d\hat{c}^{+}_{\sigma}\hat{c}_{\sigma}
+ U\hat{n}_{\uparrow}\hat{n}_{\downarrow}, \nonumber \\
\hat{V}&=&\sum_{\eta, k, \sigma}(V_\eta \hat{a}^+_{\eta, k,
\sigma} \hat{c}_{\sigma} + h.c.), \label{eqn:Hamiltonian}
\end{eqnarray}
\noindent where $\eta\in(L, R_1, R_2)$ denotes the lead label, $k$
is the lead electron momentum, $\sigma$ is the electron spin,
$\hat{a}_{\eta, k, \sigma}^+(\hat{a}_{\eta, k, \sigma})$ is the
creation (annihilation) operator for the lead electrons,
$\hat{c}_{\sigma}^+(\hat{c}_{\sigma})$ is the creation
(annihilation) operator for the dot electrons, $\hat{n}_{\sigma}$
is the number operator for the dot electrons with spin $\sigma$,
and $U$ denotes the charging energy. The tunneling matrix element
between the dot and the lead states is given by $V_\eta$.

As shown in Fig.\ref{fig:DotSystem}(a), two electrons co-tunnel
through the dot and become spatially separated at the output leads
$R_1$ and $R_2$. Any single-electron tunneling is forbidden in
this model. This can be achieved by biasing the dot system such
that no energy of a single left lead electron equals the energy of
a right lead electron, i.e.,
$\varepsilon_{L,k}\neq\varepsilon_{R_1, k_1}\neq\varepsilon_{R_2,
k_2}$, and that the conservation of energy holds for virtual
two-electron co-tunneling,
$\varepsilon_{L,k}+\varepsilon_{L,k^{'}}=\varepsilon_{R_1,
k_1}+\varepsilon_{R_2,k_2}$. Through out the paper, we assign the
dot level energy $\varepsilon_d$ to be zero without loss of
generality. The energy separation for the left lead electrons is
required to be smaller than the energy separation for the right
leads, i.e., $\triangle_L<\triangle_R$. On the other hand, the
widths of the lead energy bands are required to be narrow so that
the leads act as energy filters. The energies of the two right
leads are required to below their quasi-Fermi levels such that the
right leads are empty. In general, the left lead is assumed to be
filled of electrons. As we will show later, a net amplitude for
the singlet-spin state at the output leads is filtered out from a
two-electron input in the case that the dot is occupied by two
electrons. However, several final states may have nonzero
transition amplitudes when the dot is occupied by a single
electron.

From the Hamiltonian given in Eq.(\ref{eqn:Hamiltonian}), the
lowest order contribution to the co-tunneling processes is ${\cal
O}(V^4)$. We calculate the transition amplitudes for the
co-tunneling processes by using the transition matrix formalism,
\begin{equation}
\hat{T}(\varepsilon_i)=\hat{V}{1 \over
\varepsilon_i-\hat{H}_0}\hat{V}{1 \over
\varepsilon_i-\hat{H}_0}\hat{V}{1 \over
\varepsilon_i-\hat{H}_0}\hat{V}, \label{eqn:Transition}
\end{equation}
\noindent where $\varepsilon_i$ is the energy of the initial state
of the system and $\hat{T}(\varepsilon_i)$ is the transition
operator that is relevant to the two-electron co-tunneling. Thus
the transition amplitude between the initial state $|\varphi_i>$
and the final state $|\varphi_f>$ is given by
$<\varphi_f|\hat{T}(\varepsilon_i)|\varphi_i>$.

\section{Quantum dot occupied by two electrons}
In this section, we assume that the quantum dot is occupied by two
electrons with different spins, i.e., spin up $(\uparrow)$ and
spin down $(\downarrow)$. In general, the spins of the input
electrons will be one of the following four combinations: $(\sigma
\sigma')= (\uparrow\downarrow)$, $(\downarrow\uparrow)$,
$(\uparrow\uparrow)$ and $(\downarrow\downarrow)$. The initial
four-electron state is given by
\begin{equation}
|\varphi_i>=\hat{c}^+_{(\downarrow)}\hat{c}^+_{(\uparrow)}
\hat{a}^+_{L,k,\sigma}\hat{a}^+_{L,k',\sigma'}|\mbox{ }0>,
\label{eqn:InitialState}
\end{equation}
\noindent where $|\mbox{ }0>$ denotes the zero-particle state of
the model system.  The energies for the input electrons are
denoted as $\varepsilon_{L,k}=E_L+\triangle_L$ and
$\varepsilon_{L,k'}=E_L-\triangle_L$ such that the energy of the
initial state is $\varepsilon_i=2E_L+U$. Therefore, the
co-tunneling amplitude for $(\sigma\sigma')=(\uparrow\downarrow)$
can be obtained from the amplitude for $(\sigma
\sigma')=(\downarrow\uparrow)$ by replacing $\triangle_L$ with
$-\triangle_L$ in the amplitude. On the other hand, since the
tunneling matrix element $V_\eta$ is spin-blind the transition
amplitude for the case $(\sigma \sigma')=(\uparrow\uparrow)$
should be equal to the amplitude of the case $(\sigma
\sigma')=(\downarrow\downarrow)$. This means, only two
co-tunneling processes are needed to be calculated.

\begin{figure}
\vspace*{1.5in} \includegraphics{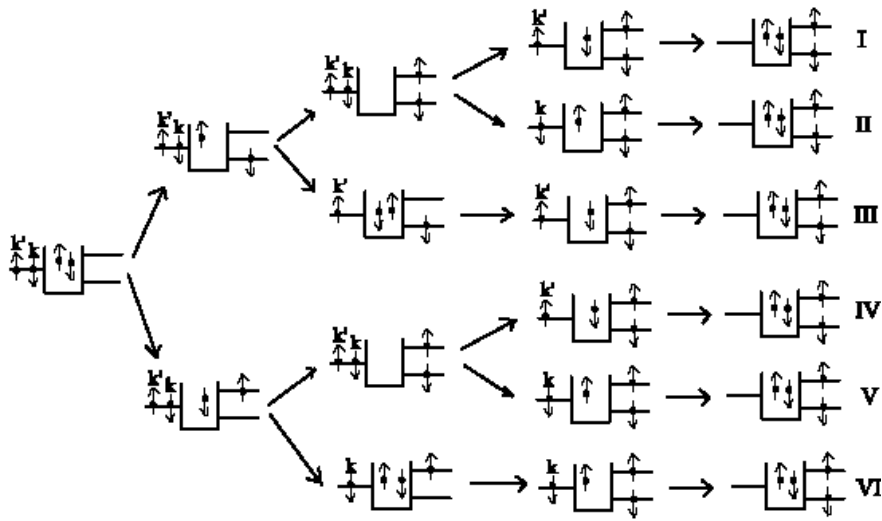}\vspace{2.0in} \caption{Multiple
paths by which two input electrons of different spins virtually
co-tunnel through a quantum dot that is occupied by two electrons
to two output leads.}\label{fig:PathC}
\end{figure}

We consider first the case $(\sigma\sigma')=(\downarrow\uparrow)$.
In this case, we cannot have two same-spin electrons at the output
leads. Since if there exist two same-spin electrons at the output
leads, then two same-spin electrons must co-exist within the
quantum dot, thus violate Pauli's exclusive principle. Therefore,
only electrons of different spins can be obtained at the output
leads. From Eq.(\ref{eqn:Transition}), the time-ordering operator
leads to twelve virtual paths by which the input electrons can
co-tunnel through the dot into output leads. Six of the twelve
paths that have "direct" time ordering are shown in
Fig.\ref{fig:PathC}. The remaining six virtual paths that have
"exchange" time ordering are obtained simply by exchanging the
output leads $R_1$ and $R_2$. In general, the portion of the
transition amplitude from the "exchange paths" acquires a relative
minus sign and can be easily obtained by simply replacing
$\triangle_R$ with $-\triangle_R$ in the amplitude from the
"direct paths". After all, the transition amplitudes for the
singlet $|\mbox{ }s>$ and triplet $|\mbox{ }t>$ final states are
\begin{eqnarray}
|\mbox{ }s>, |\mbox{ }t>&\equiv& {1 \over \sqrt{2}}
\hat{c}^+_{(\uparrow)} \hat{c}^+_{(\downarrow)}
(\hat{a}^+_{R_1(\uparrow)}\hat{a}^+_{R_2(\downarrow)}
\mp\hat{a}^+_{R_1 (\downarrow)}\hat{a}^+_{R_2(\uparrow)})|\mbox{
}0> \nonumber
\\
<t|\hat{T}(\varepsilon_i)|\varphi_i>&=&0 \nonumber \\
<s|\hat{T}(\varepsilon_i)|\varphi_i>&=&{ 2 \sqrt{2}U(E_L-U)
V^{*2}_LV_{R_1}V_{R_2}\over
(2E_L-U)((E_L-U)^2-\triangle_L^2)((E_L-U)^2-\triangle_R^2)}.
\label{eqn:ampC}
\end{eqnarray}
\noindent The triplet transition amplitude destructively
interfere, similar to the case of co-tunneling through an empty
dot \cite{Oliver}.  The singlet transition amplitude is also
destructive, but not complete when the on-site Coulomb interaction
$U\neq 0$. However, as different to the case of co-tunneling an
empty quantum dot, all virtual paths in Fig.\ref{fig:PathC} are
$U$-dependent thus are suppressed as $U\rightarrow\infty$.

The remaining cases to be considered are
$(\sigma\sigma')=(\uparrow\uparrow)$ and $(\downarrow\downarrow)$.
In these cases, the output electrons at the two right leads always
have the same spins as those of the input electrons. Take the case
$(\sigma\sigma')=(\uparrow\uparrow)$ as an example, the final
state of the dot system must be $\hat{c}^+_{(\downarrow)}
\hat{c}^+_{(\uparrow)}\hat{a}^+_{R_1
(\uparrow)}\hat{a}^+_{R_2(\uparrow)})|\mbox{ }0>$. The
time-ordering operator in Eq.(\ref{eqn:Transition}) thus leads to
four unique time orderings. We find that these four virtual paths
destructively interfere and give vanishing amplitude, independent
of the on-site Coulomb energy $U$.

Base on the above calculation, we now conclude that only net
singlet transition amplitude can be obtained in the case of two
electrons occupying the quantum dot. The left lead can be a Fermi
metal which is filled of electrons such that the input electrons
can be an arbitrary selection of two electrons of various types of
spins. Any non-vanishing co-tunneling of two electrons from the
left lead thus generates entangled spin-singlet electrons at the
output leads.

\section{Quantum dot occupied by a single electron}
In this section, we assume that the quantum dot is occupied by a
single electron with definite spin $\sigma$. Without loss of
generality, we assume the spin of the dot electron to be spin-down
$\sigma=(\downarrow)$.  Again, there are four possible
combinations for the spins of the input electrons,
$(\sigma\sigma')=(\uparrow\downarrow)$, $(\downarrow\uparrow)$,
$(\uparrow\uparrow)$ and $(\downarrow\downarrow)$. The first two
combinations are actually the same under the transformation
$\triangle_L \rightarrow -\triangle_L$ except for an extra factor
$(-1)$ which is required due to the exchange of operators for
input electrons. The last two cases are same-spin electrons
co-tunneling through the dot. However, these two cases are
intrinsically different from each other since one of them must
have different spin as that of the dot electron.

\begin{figure}
\vspace*{1.5in} \includegraphics{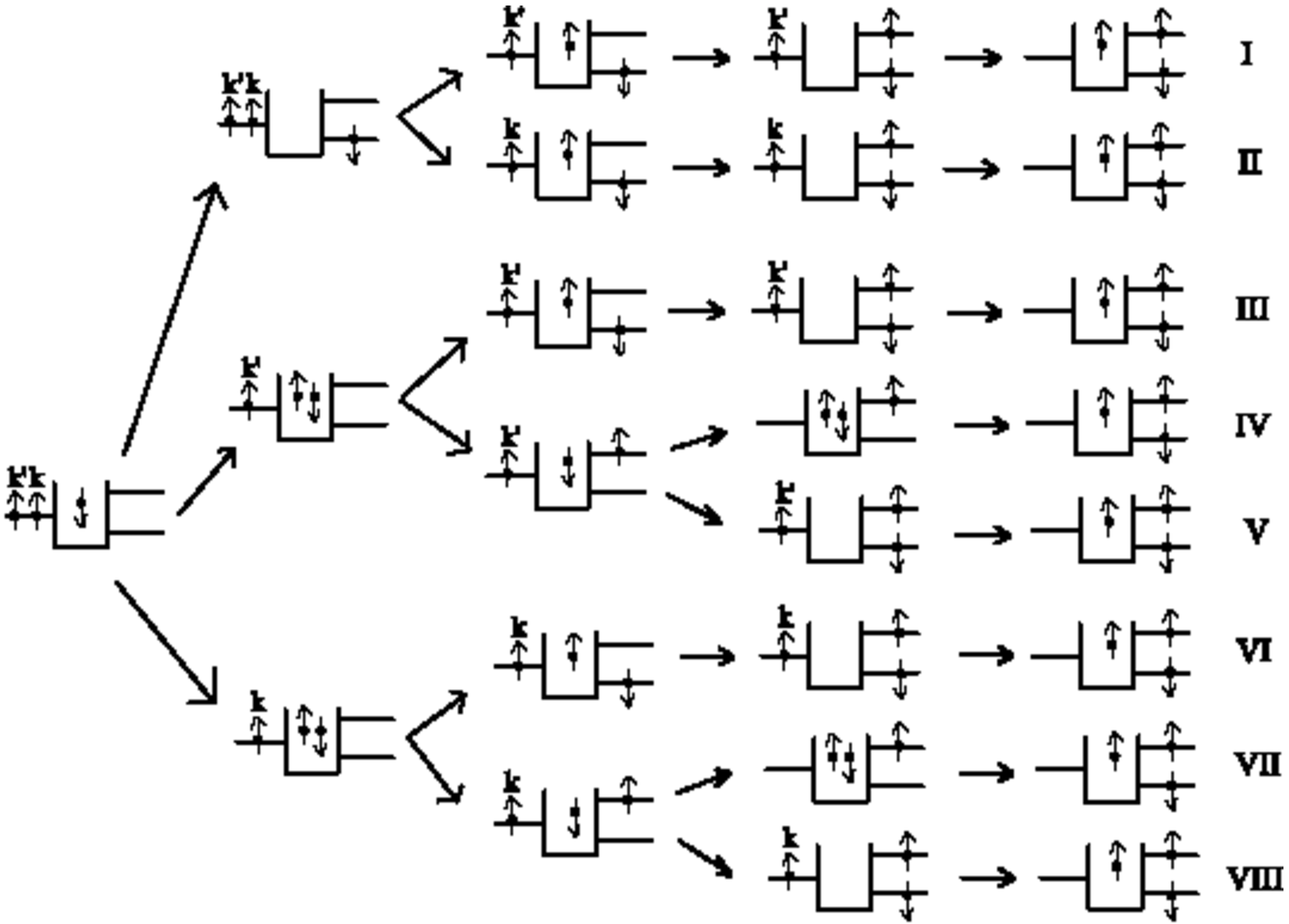}\vspace{2.0in} \caption{Multiple
paths by which two same-spin electrons virtually co-tunnel through
a quantum dot that is occupied by a single electron to two output
leads.} \label{fig:PathB}
\end{figure}

We consider first the cases of same-spin co-tunneling. For the
case of $(\sigma\sigma')=(\uparrow\uparrow)$, the initial state of
the quantum dot system is
$|\varphi_i(\uparrow\uparrow)>\equiv\hat{c}^+_{(\downarrow)}\hat{a}^
+_{L,k,(\uparrow)} \hat{a}^+_{L,k',(\uparrow)}|\mbox{ }0>$, i.e.,
both input electrons have different spins than that of the dot
electron. The output electrons at the two right leads may form a
spin-triplet, a spin-singlet, or a same-spin $(\uparrow\uparrow)$
state which corresponds to a spin-up, a spin-up, or a spin-down
dot electron confined within the dot, respectively.  Therefore,
the possible final states of the dot system are $|\uparrow, s>,
|\uparrow, t>\equiv {1 \over \sqrt{2}} \hat{c}^+_{(\uparrow)}
(\hat{a}^+_{R_1(\uparrow)}\hat{a}^+_{R_2(\downarrow)}
\mp\hat{a}^+_{R_1 (\downarrow)}\hat{a}^+_{R_2(\uparrow)})|\mbox{ }
0>$, and $ |\downarrow, \uparrow\uparrow>\equiv
\hat{c}^+_{(\downarrow)} \hat{a}^+_{R_1(\uparrow)}\hat{a}^+
_{R_2(\uparrow)}|\mbox{ }0>$. For the case that a singlet or a
triplet state is generated at the output leads, there are sixteen
virtual paths by which the input electrons can virtually co-tunnel
through the dot to the right leads. Among these virtual paths,
eight out of the sixteen paths that correspond to "direct" time
ordering are shown in Fig.\ref{fig:PathB}. The remaining eight
paths that have "exchange" time ordering are due to the exchange
of the output leads $R_1$ and $R_2$. On the other hand, there are
four unique virtual co-tunneling paths that contribute to the
transition amplitude of a same-spin final state $|\downarrow,
\uparrow\uparrow>$ at the output leads. Two of them can be viewed
as having "direct" time ordering, and the remaining two virtual
paths correspond to "exchange" time ordering. For the case of
$(\sigma\sigma')=(\downarrow \downarrow)$, the initial state of
the dot system is
$|\varphi_i(\downarrow\downarrow)>\equiv\hat{c}^+_{(\downarrow)}
\hat{a}^+_{L,k,(\downarrow)} \hat{a}^+_{L,k',(\downarrow)}|\mbox{
} 0>$, i.e., all input and dot electrons have the same spin.
Therefore there is only one possible final state for the dot
system, $|\downarrow,\downarrow \downarrow>\equiv
\hat{c}^+_{(\downarrow)} \hat{a}^+_{R_1(\downarrow)}\hat{a}^
+_{R_2(\downarrow)}|\mbox{ }0>$. Again, there are four paths that
electrons can virtually co-tunnel through the quantum dot to the
output leads. Two of the paths can be viewed as having "direct"
time orderings, and the other two paths have "exchange" time
orderings. However, all these four paths are not sensitive to the
on-site Coulomb energy since no same-spin electrons can occupy the
same energy level of the quantum dot. After summing all
contributions from the virtual paths, we get the following
transition amplitudes,
\begin{eqnarray}
<\uparrow,s|\hat{T}(\varepsilon_i)|\varphi_i(\uparrow\uparrow)>&=&
{-2\sqrt{2}\triangle_L U^2V_{L}^{*2}V_{R_1}V_{R_2} \over
(E_L^2-\triangle_L^2)(\triangle_L^2-\triangle_R^2)
((E_L-U)^2-\triangle_L^2)} \nonumber \\ <\uparrow, t
|\hat{T}(\varepsilon_i)|\varphi_i(\uparrow\uparrow)>&=&0 \nonumber
\\ <\downarrow,\uparrow\uparrow|\hat{T}(\varepsilon_i)
|\varphi_i(\uparrow\uparrow)>&=&0 \nonumber \\<\downarrow,
\downarrow \downarrow||\hat{T}(\varepsilon_i)
|\varphi_i(\downarrow\downarrow)>&=&0.
\label{eqn:sSpinOneDot}
\end{eqnarray}
\noindent As a result, only the spin-singlet final state can get
nonzero amplitude after same-spin input electrons co-tunneling
through the dot to two output leads. This implies that if only the
spin-polarized electrons with spins opposite to the dot electron
spin are prepared and then injected from the input lead to the
dot, a spin-singlet final state is always generated at the two
output leads. Recent developments on spin-injection technology
\cite{spininjection} may help in realizing this spin-entangling
mechanism. Again, turning off of the on-site Coulomb energy will
lead to vanishing spin-singlet amplitude. Similar to the case in
Ref.\cite{Oliver}, the increase of $U$ will not lead to vanishing
singlet amplitude. It simply suppresses the contributions from all
the virtual co-tunneling paths but the paths $(\mbox{I},
\mbox{II})$ in Fig.\ref{fig:PathB}.

\begin{figure}
\includegraphics{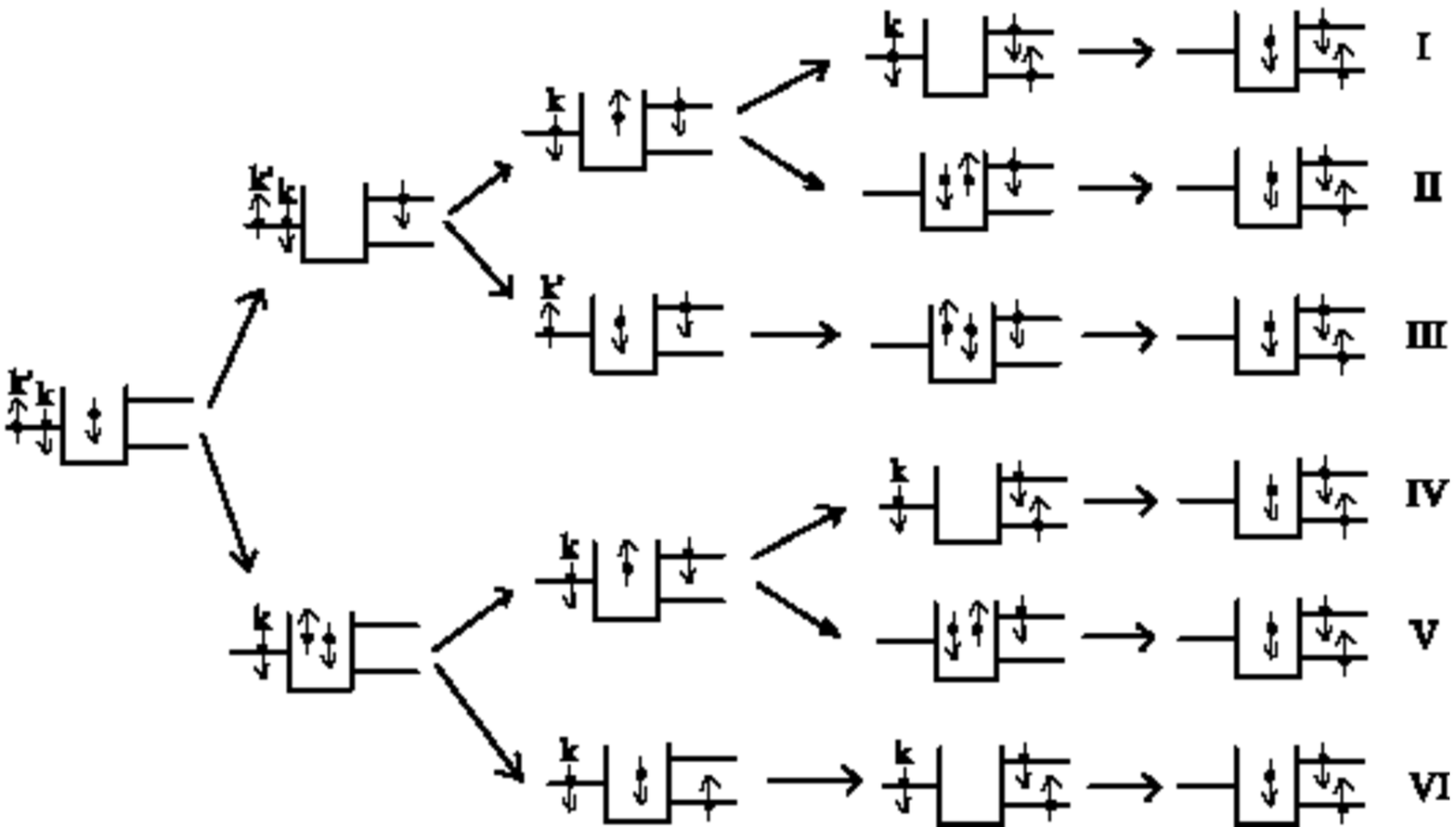}\vspace{1.8in} \caption{Multiple paths by
which two different-spin electrons virtually co-tunnel through a
quantum dot that is occupied by a single electron to two output
leads.} \label{fig:PathA}
\end{figure}

Next we consider the case that $(\sigma \sigma')=(\downarrow
\uparrow)$. In this case, the initial state of the quantum dot
system is $\hat{c}^+_{(\downarrow)}\hat{a}^+_{L,k,(\downarrow)}
\hat{a}^+_{L,k',(\uparrow)}|\mbox{ }0>$. The output electrons at
the two right leads may form a spin-triplet, a spin-singlet, or a
same-spin $(\downarrow\downarrow)$ state which corresponds to a
spin-down, a spin-down, or a spin-up dot electron confined within
the dot, respectively. Therefore, the possible final states of the
dot system are $|\downarrow,s>, |\downarrow,t>\equiv {1 \over
\sqrt{2}} \hat{c}^+_{(\downarrow)}
(\hat{a}^+_{R_1(\uparrow)}\hat{a}^+_{R_2(\downarrow)}
\mp\hat{a}^+_{R_1 (\downarrow)}\hat{a}^+_{R_2(\uparrow)})|\mbox{ }
0>$, and $|\uparrow,\downarrow\downarrow> \equiv\hat{c}^+
_{(\uparrow)} \hat{a}^+_{R_1(\downarrow)}\hat{a}^
+_{R_2(\downarrow)}|\mbox{ }0>$. In this case, there are twelve
virtual paths that contribute to the transition amplitudes for the
spin-singlet or the spin-triplet final state. Six out of the
twelve paths which correspond to "direct" time ordering are shown
in Fig.\ref{fig:PathA}. The remaining six paths then correspond to
"exchange" time ordering. On the other hand, there are eight
virtual paths contributing to the transition amplitude for
$|\uparrow,\downarrow\downarrow>$ final state. After summing all
virtual paths, we obtain the following transition amplitudes:
\begin{eqnarray}
{<\downarrow,s|\hat{T}(\varepsilon_i)|\varphi_i> \over
V_L^{*2}V_{R_1}V_{R_2}}&=&
{\sqrt{2}U\{(E_LU-E_L^2+\triangle_R^2)(\triangle_L^2
-\triangle_R^2
+\triangle_LU)+U^2\triangle_R^2\}\over(E_L^2-\triangle_R^2)(\triangle_L^2
-\triangle_R^2)((E_L-U)^2-\triangle_R^2)(E_L+\triangle_L)(E_L-\triangle_L-U)
} \nonumber \\ {<\downarrow,t|\hat{T}(\varepsilon_i)|\varphi_i>
\over V_L^{*2}V_{R_1}V_{R_2}}&=&{-\sqrt{2}\triangle_R U^2 \over
(E_L^2-\triangle_R^2)(\triangle_L^2-\triangle_R^2)((E_L-U)^2-\triangle_R^2)}
\nonumber \\
<\uparrow,\downarrow\downarrow|\hat{T}(\varepsilon_i)|\varphi_i>&=&
- <\downarrow,t|\hat{T} (\varepsilon_i)|\varphi_i>.
\end{eqnarray}
\noindent The above result shows that net amplitudes may exist for
all the final states $|\downarrow,s>$, $|\downarrow,t>$ and
$|\uparrow,\downarrow\downarrow>$ if $U\neq 0$. All amplitudes
vanish when $U=0$. However, there is a chance that the transition
amplitude for $|\downarrow,s>$ final state will vanish:
\begin{equation}
(E_LU-E_L^2+\triangle_R^2)(\triangle_L^2 -\triangle_R^2
+\triangle_LU)+U^2\triangle_R^2 =0. \label{eqn:TuneoffS}
\end{equation}
\noindent For example, while $E_L <0$ and $\triangle_L
<\triangle_R$ are presumed in this paper, the above condition
leads to $\triangle_R^2\simeq|E_L|\triangle_L$ in the
$U\rightarrow\infty$ limit.  In this situation only the
spin-triplet $|\downarrow,t>$ and the same-spin state
$|\uparrow,\downarrow \downarrow>$ are allowed as the final states
for the dot system.

Based on the above results, we find that the dot system may
generate each of the four possible final states including
$|\downarrow, s>$, $|\downarrow, t>$, $|\uparrow,s>$ and
$|\uparrow,\downarrow\downarrow>$. When the on-site Coulomb energy
is turned off, the transition amplitudes of all these final states
also vanish. Otherwise, the last three final states always have
non-zero transition amplitudes no matter what the model parameters
are chosen. The transition amplitude for $|\downarrow, s>$ may be
tuned to zero by satisfying Eq.\ref{eqn:TuneoffS}. This implies
that once the final dot electron is detected to have the same spin
as that of the initial dot electron, we can fully infer that a
pair of entangled electrons which form a spin-triplet state are
generated at the two output leads.

\section{Conclusion}
In this paper, we discussed the possibility of using non-empty
quantum dots as the spin entanglers. The dot system consists of a
single quantum dot which has only a single energy level within the
dot, one input lead which is filled of electrons, and two output
leads which are empty. Two situations are discussed in this paper
: the dot system occupied by one, or by two dot electrons within
the quantum dot. The dot system is arranged such that any
single-electron tunneling process is suppressed due to energy
conservation. Only two input electrons can co-tunnel through the
dot to the output leads.

When the dot is occupied by two dot electrons, we found that the
dot system can be a good electron entangler which always filters
the singlet-state portion of the two-electron input and generates
a non-local spin-singlet state at the output leads. This result is
similar to that in Ref.\cite{Oliver} which considers an empty
quantum dot as the electron entangler. In both systems, the
on-site Coulomb interaction $U$ acts as a nonlinear mediator which
mediates electron entanglement. When $U=0$ all transition
amplitudes vanish. When the on-site Coulomb energy is turned on,
both systems generate only spin-singlet states at the output
leads.

When the dot is occupied by a single dot electron with definite
spin (assumed spin-down in the paper), we found that the dot
system no longer filters any particular portion of the
two-electron input when the injected electrons are unpolarized. In
general, there are nonzero transition amplitudes for four
different final states: $|\downarrow,s>$, $|\downarrow,t>$,
$|\uparrow, s>$, and $|\uparrow, \downarrow\downarrow>$. Again,
all these four amplitudes vanish when the Coulomb energy $U$ is
turned off. However, when Eq.(\ref{eqn:TuneoffS}) is satisfied the
amplitude for the final state $|\downarrow, s>$ can be tuned to
zero. If this is the case, only the triplet final state
$|\downarrow,t>$ will not flip the spin of its dot electron after
co-tunneling. This means that when the dot electron is detected to
remain in the same spin state, a pair of entangled electrons which
form a spin-triplet must be generated at the output leads. On the
other hand, the dot system will filter out spin-singlet final
state at the two output leads if only polarized input electrons
with spins opposite to the dot-electron spin are prepared and
injected from the input lead.

\begin{acknowledgments}
We thank Institute of Physics in Academia Sinica for helpful
supports during my visit to the institute. This work was also
supported in part by National Science Council of Taiwan
NSC90-2112-M-033-011.
\end{acknowledgments}

%\bibliography{}%

\newpage


\end{document}